\documentclass[pra,twocolumn,preprintnumbers]{revtex4}
\usepackage{graphicx}

\begin{document}
\preprint{ITEP-TH-48/09}

\title{Information Theory and Renormalization Group Flows}

\author{S. M. Apenko }
\affiliation{ I E Tamm Theory Department, P N Lebedev Physics Institute, Moscow,
119991, Russia} \email{apenko@lpi.ru}

\begin{abstract}
We present a possible approach to the study of the renormalization group (RG) flow
based entirely on the information theory. The average information loss under a single
step of Wilsonian RG transformation is evaluated as a conditional entropy of the fast
variables, which are integrated out, when the slow ones are held fixed. Its positivity
results in the monotonic decrease of the informational entropy under renormalization.
This, however, does not necessarily imply the irreversibility of the RG flow, because
entropy is an extensive quantity and explicitly depends on the total number of degrees
of freedom, which is reduced. Only some size-independent additive part of the entropy
could possibly provide the required Lyapunov function.  We also introduce a mutual
information of fast and slow variables as probably a more adequate quantity to
represent the changes in the system under renormalization and evaluate it for some
simple systems. It is shown that for certain real space decimation transformations the
positivity of the mutual information directly leads to the monotonic growth of the
entropy per lattice site along the RG flow and hence to its irreversibility.
\end{abstract}

\maketitle

\section{Introduction}
Renormalization group (RG), which is a powerful instrument for analyzing different
strongly coupled systems \cite{W}(see also \cite{BBD} for more recent reviews), is
usually based upon an appropriate division of the whole set of variables into two
subsets: so called ``fast'' and ``slow'' variables with subsequent elimination of the
fast ones. The resulting slow system is usually expected to behave essentially in the
same way as the initial one but with coupling constants changed. Successive
application of this transformation results in a flow in the space of couplings. After
each step of the RG transformation the information about exact values of fast
variables is lost for the observer to whom only slow variables are available. It seems
natural then to expect, that this information loss should lead to the irreversible
character of the corresponding RG flow.

This irreversibility is a subject of an intensive study for several decades now,
starting from the famous work of Zamolodchikov \cite{Z}, who showed under certain
assumptions that in two dimensions (2D) there exists a function on a space of coupling
constants (a kind of Lyapunov function) which always decreases along the RG flow and
at fixed points coincides with the central charge $c$ of corresponding conformal
theories. ``Irreversibility'' here means that the flow in the space of couplings
resembles that of a simple dissipative system with a given trajectory never returning
back to its starting point, thus excluding limit cycles or more complex strange
attractors. In 2D theories the $c$-theorem of Ref. \cite{Z} states that the RG
evolution is exactly of this type and looks like a simple monotonic flow downhill in
the couplings' space from fixed points with large $c$ to those with smaller values of
the central charge.

Much work have been done in this direction and the $c$-theorem was studied in detail
with its possible generalizations to higher dimensions (see e.g. \cite{c} and
references therein and also \cite{Ftheor} for quite recent development). Still the
interesting possibility that in general the flow may be more complicated, and may even
exhibit chaotic behavior, like in case of many other nonlinear mappings, attracts some
attention \cite{M} (probably the first example of a chaotic RG flow was found for spin
systems on hierarchical lattices \cite{h}, see also \cite{Da,D}). The models where
such peculiarities are observed are however rather artificial, so that the question
whether they could take place in more realistic cases remains somewhat unclear.

Though this problem is rather complicated mathematically it looks much simpler from
the physical point of view. The Wilsonian RG transformation is designed in such a way
that all finite dimensionless correlation lengths in a translationally invariant
system always decrease \cite{W}. To say this another way, all masses (measured in
units of an ultra-violet cut-off) grow and it seems quite natural that the effective
number of massless modes, (in 2D measured by the central charge $c$) cannot increase.
The argument based on the growth of masses, though not completely rigorous, is quite
general, hence something unusual, like limit cycles in the flow, may be expected only
in some special systems with only infinite correlation lengths or with infinite number
of correlation lengths which can be arbitrarily large (when we have e.g. a
self-similar spectrum of masses converging to zero \cite{GW}). But in systems with
finite number of well defined correlation lengths the RG flow should probably always
be irreversible.

This argument however seems to be unrelated to the aforementioned information loss and
information aspects probably play no role in the irreversibility, which may seem
somewhat strange. Probably the most known attempt to study RG irreversibility using
information theory tools is the approach of Ref. \cite{G1,G2} where the relative
entropy was introduced in quantum field theory and some monotonicity theorems for it,
as a function of relevant couplings, were proved and its relation to $c$-theorem was
discussed.

In this paper we present a somewhat different, entirely information theory based
approach to the problem of the RG flow, oriented more on discrete real space RG
transformations in different lattice spin systems. Clearly, information-theoretic
considerations are too general to lead to some non-trivial concrete results, but
still, as we shall see below they may impose some restrictions on the possible
character of the flow. First, in Section 2 we evaluate an exact amount of information
loss for a single step of the renormalization as a kind of conditional entropy of fast
variables (compare with \cite{L}). As one might expect, this information loss is equal
to the decrease of entropy under the RG transformation. Thus it is possible to
establish a general theorem about the monotonic decrease of the total entropy (and
relative entropy as well) under general coarse graining as a result of information
loss.

The monotonic decrease of entropy along the RG flow is neither something unusual (see
e.g. \cite{Ro}) nor very interesting, because the entropy is proportional to the total
number of degrees of freedom and should certainly decrease when some of them are
integrated out. Hence this monotonic behavior does not directly lead to the
irreversibility of the flow. An interesting thing may happen however, if the entropy
of a finite system possesses an additive size-independent part. If this part of the
entropy also decreases, as the total entropy does, then it may provide the proper
Lyapunov function, because it does not explicitly depend on the number of variables
and its change is entirely due to the flow of coupling constants. We will show that
this takes place in 1D Ising model, where this subextensive part of the entropy is
known as the ``excess entropy'' and was studied in detail as a measure of statistical
complexity of spatial structures \cite{FC}. This excess entropy equals to the mutual
entropy of two halves of the system which is actually the reason why it decreases
under coarse graining. Unfortunately we are unable to present a general theorem
concerning the monotonicity of such excess entropy in higher dimensions.

It appears possible though to prove quite a different monotonicity theorem, but only
for a certain class of real space lattice RG transformations, namely for decimations,
when exactly half of variables are eliminated on each step. The proof is based on the
analysis of a new quantity introduced in Section 3,---mutual information of slow and
fast variables. Contrary to the information loss this mutual information shows how
much information about eliminated fast variables is still present after a single step
of the Wilsonian renormalization. The nonnegativity of the mutual information may
alone impose some restrictions on the character of the flow leading to its
irreversibility.

For the decimations of the aforementioned type in lattice models we will show that the
entropy per lattice site monotonically grows as a result of the positivity of the
mutual information of spins on two identical sublattices, one of which is eliminated
in the course of the RG transformation. Therefore such decimation transformations,
though they results in highly non-linear mappings, always lead to irreversible flows
in the space of couplings. Simple examples of such a behavior are discussed in Section
4, where one- and two-dimensional classical spin models and also the continuum limit
Gaussian model are considered. Note, however, that the theorem holds true only for
systems with real actions, when information theory can be applied, while e.g. it is
not valid for the Ising model with complex external magnetic field  where the
decimation RG flow is known to be chaotic \cite{D}. The theorem is also invalid for
spin models on hierarchical lattices \cite{h}, because decimations are not associated
there with decomposition of these finite and inhomogeneous lattices into two identical
sublattices.

The entropy per site growth may seem somewhat unexpected, because it takes place not
only in the trivial case, when the flow goes to the high-temperature fixed point, but
even when we start from points located in the ordered phase. RG trajectories that
start from points below the phase transition can never end in the fully disordered
trivial fixed point, since for decimation the mean magnetization is conserved along
the flow, but nevertheless the entropy per site (and hence the amount of disorder)
will always grow. This is a peculiar feature of the decimation and does not take place
for other possible renormalization schemes, like e.g. the majority rule block spin
transformation also discussed in Section 4.

\section{Information loss and the RG flow}

 Suppose that we somehow manage to divide our variables
into ``fast'' $\psi$ and ``slow'' ones $\eta$. For the Wilsonian momentum space RG the
fast variables are high Fourier harmonics of the fields, but for real space
renormalization in lattice systems $\psi$'s are just spins on some sublattice, so that
they are not actually ``fast'' in any sense. Still we will use these names to
distinguish degrees of freedom to be integrated out and those that remain. The
partition function is given by the sum over all fields, which symbolically may be
written as
\begin{equation}\label{Z}
  Z=\sum_{\psi,\eta}\exp\left(-S(\psi,\eta)\right),
\end{equation}
where $S(\psi,\eta)$ is the action (or energy divided by temperature for classical
lattice systems). The fields normally depend on all space-time coordinates, so that
the sums in Eq. (\ref{Z}) are actually path integrals, properly regularized. Now, the
first step of the RG transformation is to integrate out the fast variables $\psi$ to
obtain an effective action for the slow ones, $S'(\eta)$, defined by
\begin{equation}\label{eff}
  \exp\left(-S'(\eta)\right)=\sum_{\psi}\exp\left(-S(\psi,\eta)\right)
\end{equation}
This transformation is then repeated many times resulting in the RG flow of the
corresponding actions (or coupling constants if the form of the action is fixed).

\subsection{Information loss and entropy}

Certainly, after the fast variables are eliminated, information about their exact
values is lost. One can ask then, what is the precise amount of the information loss.
For a moment let us fix the slow variables $\eta$, then the conditional probability
for $\psi$ (normalized to unity) is
\begin{eqnarray}\label{pcond}
 & p(\psi|\eta)=\exp\left(-S(\psi,\eta)+S'(\eta)\right),\\
  &\sum_{\psi}p(\psi|\eta)=1. \nonumber
\end{eqnarray}
It seems natural then to define the information (measured in `nats') related to some
field configuration $\psi$ in the fixed background $\eta$ by the usual Shannon formula
\begin{equation}\label{i}
  I_{\psi}=-\ln p(\psi|\eta),
\end{equation}
and the information loss after the fields $\psi$ are integrated out as an average of
(\ref{i}) over the distribution (\ref{pcond})
\begin{equation}\label{ia}
  \langle I_{\psi}\rangle=-\sum_{\psi}p(\psi|\eta)\ln p(\psi|\eta)
\end{equation}
This quantity still depends on the chosen configuration of the slow fields $\eta$.
Actually $\langle I_{\psi}\rangle$ is similar, in a sense, to the Boltzmann entropy of
a `macrostate' specified by the coarse-grained variables $\eta$. Indeed, in case of
`equipartition', when $p(\psi|\eta)$ does not depend on $\psi$ in some domain of fast
variables the information loss Eq.(\ref{ia}) is just the logarithm of the number of
`microstates' corresponding to a given macrostate.

We propose then that the total information loss $\delta I$ is an average of $\langle
I_{\psi}\rangle$ over all configurations of $\eta$ with the corresponding weight
$p(\eta)=1/Z\exp(-S')$, i.e.
\begin{equation}\label{it}
  \delta I=-\sum_{\eta}p(\eta)\sum_{\psi}
  p(\psi|\eta)\ln p(\psi|\eta).
\end{equation}
This is obviously the {\it conditional entropy} $H(\psi|\eta)$ of two sets of random
variables $\psi$ and $\eta$ well known from the information theory \cite{Cov}. This is
the average entropy of the fast variables given the state of the slow subsystem. The
conditional entropy of this kind usually measures the amount of information loss in a
noisy channel when $\psi$ is an input signal and $\eta$ is an output. One can also
view the RG transformation as an action of a smoothing filter, then $\delta I$ is the
information loss due to smoothing. The conditional entropy as a measure of information
loss also have been introduced in Ref. \cite{L} where a coarse graining of polymer
configurations was considered.

The above estimate of the information loss introduces a somewhat new paradigm of
renormalization. We see now that it is possible to view the fast fields $\psi$ as an
input signal, not directly available to a receiver, while the slow field $\eta$ is an
output. After renormalization only the output signal is available and one can only try
to restore the input signal more or less accurately. This is a typical situation of a
signal transmission through a noisy channel where information loss may be estimated as
a conditional entropy (\ref{it}). But this analogy suggests that it may probably be of
more use to discuss not only the information loss but also the information that is
preserved, i.e. information about fast variables that is stored in the action for the
slow ones (see Section 3).

Substituting the probability distributions into Eq. (\ref{it}) we easily obtain
\begin{equation}\label{I}
  \delta I=H(\psi|\eta)=\langle S\rangle_S-\langle S'\rangle_{S'}=H-H',
\end{equation}
where the brackets $\langle \ldots\rangle$ mean the average over the fields with the
corresponding actions (clearly, for continuous theory some regularization is required
for these formulas to make sense), the initial informational entropy $H$ of the
distribution $p(\psi,\eta)$ is given by
\begin{equation}\label{H}
  H=-\sum_{\eta,\psi}p(\psi,\eta)\ln p(\psi,\eta)=\langle S\rangle_{S}+\ln Z,
\end{equation}
where
\begin{equation}\label{p}
  p(\eta,\psi)=\frac{1}{Z}\exp(-S(\eta,\psi))
\end{equation}
and the entropy of the slow field $H'=H(\eta)$ is given by the same formula with
$p(\eta,\psi)\rightarrow p(\eta)$ and $S\rightarrow S'$. Recall that our RG
transformation is defined in such a way that the partition function $Z$ remains
unchanged. Also note that the new action $S'$, defined by (\ref{eff}), always contains
an additive constant term, but it does not contribute to the entropy and may be
ignored in the evaluation of $H'$.

The Shannon entropy $H$ used here is the entropy of the fluctuating fields and it
measures in fact how large these fluctuations are. For a classical system $H$
coincides with its thermodynamic entropy $H_T$, but in quantum systems (in the
continuum limit) they are different, because $H_T$ is related to fluctuations in
occupation of energy levels and not to that of fields as functions of space-time
coordinates. While $H$ is dominated by an ultra-violet divergent ``bulk''
contribution, proportional to the space-time volume, the quantum thermodynamic entropy
$H_T$ looks, in a sense, as a finite-size correction to a leading term in $H$. Indeed,
for example, for a 2D system on a large cylinder of length $L$ and circumference
$\beta$ which may be viewed as a 1D quantum system at a temperature $T=1/\beta$ we
have $\ln Z\sim L\beta$ and the same is true for $H$, but this bulk term obviously
does not contribute to the thermodynamic entropy $H_T=\partial (T\ln Z)/\partial T$.
Hence non-zero $H_T$ arises only from finite size corrections to $\ln Z$ (see Section
4.5 for another example).

Thus the information loss $\delta I$ is merely the difference of the initial entropy
$H=H(\psi,\eta)$ and the entropy of the remaining slow variables $H(\eta)$. For
discrete variables $\eta$, $\psi$ (e.g. for classical spin systems) both entropies and
the conditional entropy are always nonnegative, $H\geq 0$, $H'\geq 0$ and
$H(\psi|\eta)\geq 0$ and hence
\begin{equation}\label{eg}
  H\geq H',
\end{equation}
which means that the entropy of the remaining variables decreases along the RG flow.

At first glance this decrease of entropy may seem strange, since if the flow is e.g.
directed toward the high temperature fixed point the system obviously gets more
disordered after renormalization. Recall, however, that it is the total entropy, an
extensive quantity proportional to the system size, that decreases according to Eq.
(\ref{eg}) and this decrease is mainly due to the fact that the total number of
variables is reduced after some of them are integrated out. For example, in large
lattice systems $H\simeq Nh$, where $N$ is the number of sites which decreases under
real space RG transformation. The entropy per lattice site $h$ which is a more
adequate measure of disorder may well increase along the RG flow.

Therefore the information loss $\delta I$ is not very informative about the change in
physics under renormalization and the monotonic decrease of $H$ does not in general
give rise to a proper Lyapunov function defined on the space of coupling constants.

This is not very surprising because the total information loss is a rather crude
characteristic of the RG transformation. An interesting possibility still remains,
that some part of this information loss may provide the required monotonic function.
The most natural choice probably would be some volume-independent additive part in $H$
for a finite system, something that may be called e.g. information about boundary
conditions. An example of such a function will be explicitly presented in section 4.1
for the one-dimensional Ising model. How to obtain possible extension of this result
to higher dimensions is however not clear.

To conclude this subsection it should be mentioned, that for continuous field
variables when we deal not with true probabilities but probability distributions and
sums are replaced by integrals, direct application of the above formulas may lead to
negative values of entropies \cite{Cov} and to negative $\delta I$ which certainly is
not acceptable. To avoid this one should either somehow make the variables $\eta$,
$\psi$ discrete (``digitize'' them) or better use some other quantities, which do not
suffer from such a drawback, like e.g. the {\em relative entropy}.

\subsection{Relative entropy}

Relative entropy is the basic concept of the approach to renormalization developed in
a series of papers \cite{G1,G2}. For the sake of completeness we present here a short
discussion of its properties with regard to the Wilsonian RG transformation.

Consider some statistical system with variables $\phi$, and two probability
distributions $\mu(\phi)$ and $\nu(\phi)$. Then we may introduce the relative entropy
$H_{rel}$ as the so called Kullback-Leibler distance (divergence) $D(\mu||\nu)$
between these probability distributions
\begin{equation}\label{ra}
  H_{rel}=D(\mu||\nu)=\sum_{\phi}\mu(\phi)\ln\frac{\mu(\phi)}{\nu(\phi)}
\end{equation}
This distance, though not symmetric, is nonnegative even for continuous $\phi$ and
$H_{rel}=0$ for $\mu=\nu$.

Now, the main idea of the approach \cite{G1} is to consider probability distributions
corresponding to two actions $S_1$ and $S_2$ which differ by the values of coupling
constants, and to study the relative entropy $H_{rel}=D(p_1||p_2)$ where
$p_1\sim\exp(-S_1)$ and $p_2\sim\exp(-S_2)$, and then choose $p_2$ to correspond to a
RG fixed point distribution. Thus physically this $H_{rel}$ measures how far our
system with the action $S_1$ is from a given fixed point. Then we expect the relative
entropy to behave monotonically under the change of the coupling constant,
corresponding e.g. to some relevant perturbation of the fixed point.

In fact, some rather general results are known about the behavior of $H_{rel}$ under a
coarse-graining. Let us proceed as usual and decompose the whole set of fields $\phi$
into fast $\psi$ and slow ones $\eta$. If we introduce conditional probability
$p(\psi|\eta)$ (normalized to unity $\sum_{\psi}p(\psi|\eta)=1$) and make use of
$p(\psi,\eta)=p(\eta)p(\psi|\eta)$ we can write the following chain of equalities
\begin{eqnarray}\label{ca}
H_{rel} & = & \sum_{\psi,\eta}p_1(\psi,\eta) \ln\frac{p_1(\psi,\eta)}{p_2(\psi,\eta)}=
\nonumber \\
   & = & \sum_{\psi,\eta}p_1(\eta)p_1(\psi|\eta)
   \ln\frac{p_1(\eta)p_1(\psi|\eta)}{p_2(\eta)p_2(\psi|\eta)}=
   \nonumber \\
   & = & \sum_{\eta}p_1(\eta)\ln\frac{p_1(\eta)}{p_2(\eta)}+ \nonumber\\
   & + & \sum_{\eta}p_1(\eta)\sum_{\psi}p_1(\psi|\eta)
   \ln\frac{p_1(\psi|\eta)}{p_2(\psi|\eta)}.
\end{eqnarray}
The first term in the resulting expression is obviously the relative entropy after
renormalization, $H'_{rel}$, while the last term is known in information theory as a
{\em conditional relative entropy} which is always nonnegative \cite{Cov}. Hence we
have
\begin{equation}\label{ha}
  H_{rel}\geq H'_{rel}
\end{equation}
This result, as well as the derivation of Eq. (\ref{ca}), is in fact widely known and
usually referred to as a decrease of relative entropy under any kind of
coarse-graining \cite{Cov}. The decrease of relative entropy for continuum Gaussian
model with lowering the ultra-violet cutoff \cite{G1} is just one example of this
general phenomenon.

One can also view Eq. (\ref{ca}) as a generalization of the expression (\ref{I}) for
the information loss. Then the last term in Eq. (\ref{ca}) may probably be called
``conditional information loss''. Since this is nonnegative also for continuous
variables it looks more preferable for quantum field theory, than the information loss
from the previous subsection.

Unfortunately inequality (\ref{ha}) for the relative entropy is also not very useful
because normally $H_{rel}$ is an extensive quantity and its monotonic decrease again
may be attributed to the reduction in the number of degrees of freedom (effective
shrinking of the system under renormalization) and it is a specific quantity
$h_{rel}=H_{rel}/N$, where $N$ is e.g. a number of lattice sites, which is physically
relevant. A more careful analysis shows however that $H_{rel}$ is much better that
entropy $H$ or similar quantities and possesses interesting monotonicity properties
with respect to couplings (see Refs. \cite{G1,G2} for details). For example, close to
a critical point with a diverging dimensionless correlation length $\xi$ the relative
entropy taken with respect to the fixed point distribution in many cases behaves as
$\sim N/{\xi}^d$ (for $d$-dimensional lattice system) and is universal, depending not
on the total number of variables, but only on the number of blocks of size $\xi$. This
means that relative entropy does not contain redundant ``low-level'' information. When
this is true $h_{rel}\sim 1/\xi$ is obviously monotonic along the Wilsonian RG flow,
which is a particular example of how the informational ``distance'' grows on a
trajectory moving away from the unstable fixed point.

If we take $\nu$ in (\ref{ra}) to be a uniform distribution $\nu={\rm const}$, then
$H_{rel}=-\ln\nu-H$, where $H$ is the entropy of a given distribution $\mu$.
Therefore, if the first term $\ln\nu$ remains constant, the decrease of the relative
entropy under a coarse-graining is usually related to the growth of the entropy $H$
\cite{Cov}. In our case, however, the first term also changes, because $\ln\nu\sim N$
due to normalization condition for $\nu$ and therefore $H_{rel}$ and $H$ both
decrease.

\section{Mutual information of fast and slow variables}

We now introduce another information-related quantity, the so called {\it mutual
information} of the fast and slow variables, which seems to be free of the
aforementioned drawbacks. We shall demonstrate now that the nonnegativity of this
mutual information may itself impose some restriction on the RG flow.

To define this mutual information, we need to perform a transformation opposite, in a
sense, to the usual RG transform of Eq. (\ref{eff}), namely to integrate out the slow
variables. This will result in an effective action $\tilde{S}'$ for the fast variables
$\psi$, their probability distribution being $p(\psi)\sim \exp(-\tilde{S}'(\psi))$,
and the corresponding entropy will be denoted by $H(\psi)$. Then the mutual
information of two sets of variables $\psi$ and $\eta$ is defined as
\begin{equation}\label{mud}
  I=\sum_{\psi,\eta}p(\eta,\psi)
  \ln\left[\frac{p(\eta,\psi)}{p(\eta)p(\psi)}\right]
\end{equation}
where $p(\eta,\psi)=1/Z\exp(-S)$ is the probability distribution for the initial
system (joint probability for $\eta$ and $\psi$ variables).

The mutual information is certainly a kind of relative entropy which  in our case
measures the ``distance'' $D(p(\psi,\eta)||p(\psi)p(\eta))$ between the original
system and some hypothetic one where fast and slow variables are independent, but
described by their true corresponding effective actions, i.e. $\mu=p(\psi,\eta)$ and
$\nu=p(\psi)p(\eta)$ in Eq. (\ref{ra}).

Substituting expressions for probabilities in terms of actions one can easily derive
the well known formulas from elementary information theory
\begin{eqnarray}\label{mu}
  I=H(\eta)-H(\eta|\psi)=H(\psi)-H(\psi|\eta)= \nonumber\\
  =[H(\eta)+H(\psi)]-H(\psi,\eta)
\end{eqnarray}
Here $H(\psi|\eta)=\delta I$ is the information loss after the normal RG
transformation and $H(\eta|\psi)$ is the information loss after elimination of the
slow variables, or if we adopt the view that renormalization is the information
transmission through the noisy channel $\psi\rightarrow\eta$, then $H(\eta|\psi)$ is a
measure of the noise \cite{Cov}.

In bipartite systems the mutual information is nonnegative and is known to measure the
information about one subsystem stored in the other \cite{Cov} (for independent
subsystems $I=0$), i.e. in our case it shows what amount of information about the fast
fields $\psi$ is still contained in the effective action $S'(\eta)$. Since usually RG
transformation results in the change of coupling constants it seems natural to think
that the information about the eliminated fast variables is now hidden in the new
values of these constants. Therefore $I$ should be directly related to the RG flow of
the coupling constants.

Note also that $I$ is represented in the last equation of Eq. (\ref{mu}) as a
difference of two quantities with roughly the same size dependence, so that it should
not be so sensitive to the reduction of degrees of freedom under the RG
transformation.

While in general $I$ is more difficult to calculate than the entropy, there exists an
interesting situation when $H(\psi)=H(\eta)=H'$ and $I=2H'-H$. This happens e.g. for
some decimation transformations in lattice systems, when the system is divided into
two identical sub-lattices (see next Section) and exactly half of variables (those on
one sub-lattice) are integrated out. For large enough system with $N$ sites we may
write $H=Nh$ and $H'=(N/2)h'$ where $h$ and $h'$ are the entropies per lattice site
before and after the renormalization. Then
\begin{equation}\label{mudc}
  I=N(h'-h)\geq 0\qquad\Rightarrow\qquad h'\geq h
\end{equation}
and hence the condition $I\geq 0$ directly leads to the growth of the entropy per
lattice site along the RG trajectory.

This means that such decimations are always irreversible, but suggests that
corresponding RG flows do not have fixed points for finite generic values of
couplings. Indeed, if there exists a fixed point then at this point $h=h'$ and hence
the bulk mutual information $I$ is zero. But this means that spins on two sub-lattices
are essentially uncorrelated and remain uncorrelated under renormalization, which
seems impossible for realistic spin systems with reasonable finite interactions. Note,
that this does not imply that decimation does not have nontrivial fixed points at all,
but such fixed points (apart from the trivial high-temperature one) should probably
lie at the ``boundary'' of the parameter space, when some couplings are set to
infinity, as one can see e.g. in the classical spin-1 model on a line \cite{spo} (see
also section 4.2).

The absence of a nontrivial fixed point corresponding to the phase transition for pure
decimation in 2D Ising model have been noticed already in \cite{W,KH} and was related
to the fact that decimation does not eliminates only ``short range'' correlations but
inevitably sums over some long-range behavior \cite{SB}, i.e. it is actually not a
good RG transformation.

Let us now take a look at some simple examples.

\section{Examples of the informational approach}

\subsection{One-dimensional Ising model.}

\begin{figure}[h]
\includegraphics[width=2.5in]{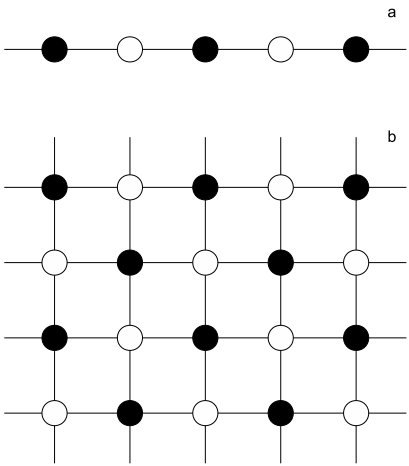}
\caption{Decimation transformations on (a) one-dimensional lattice and (b)
two-dimensional square lattice. Spins on white lattice sites are summed away.}
\label{fig1}
\end{figure}

The classical one-dimensional ferromagnetic Ising model is the most known example of
the RG transformation which here can be performed exactly. The model is defined on a
lattice with $N$ sites and the initial action is given by $S=-K\sum_i s_{i}s_{i+1}$,
where $s_i=\pm 1$. The real space RG procedure (decimation) is defined as follows: we
decompose the lattice into two sub-lattices and sum over spins on one sublattice
\cite{KH} (see Fig. 1a). The resulting model is again of the Ising type with
$K\rightarrow K'$, with twice the initial lattice spacing and $N/2$ sites. The entropy
$H$ is here the true entropy of the Ising model and the information loss, defined by
Eq. (\ref{I}), is
\begin{equation}\label{ii}
  \delta I=Nh(K)-\frac{N}{2}h(K'),
\end{equation}
where $h(K)$ is the entropy per lattice site
\begin{equation}\label{h}
  h(K)=-K\tanh K+\ln(2\cosh K),
\end{equation}
and
\begin{equation}\label{k1}
K'=(1/2)\ln(\cosh(2K)).
\end{equation}
The RG flow is very simple since the running coupling constant $K$ flows from the zero
temperature fixed point $K=\infty$ to the high temperature one at $K=0$.

Though $K'$ decreases during the RG procedure and $h(K')>h(K)$ it is easy to verify
that $\delta I$ is always positive, has its maximum $(N/2)\ln2$ at $K=0$ and then
monotonically decreases to zero as $K\rightarrow\infty$.

 It is interesting that for large $K,K'$ the information loss
can be obtained independently, so that Eq. (\ref{ii}) can be used to derive $K'$.
Indeed, if $K\rightarrow \infty$ most probable spin configurations are that of well
separated kinks (domain walls). If a spin, which is eliminated, falls inside a domain,
no information is lost, because its value does not fluctuate at large $K$ and is
completely determined by adjacent spins. But a spin that lies on a domain wall is
completely unconstrained and hence the information loss is $\ln2$. Since there are on
the average $N\exp(-2K)$ kinks at $K\rightarrow\infty$ we conclude that
\begin{equation}\label{ik}
  \delta I\simeq N\exp(-2K)\ln2+\ldots,
\end{equation}
where dots indicate terms of higher order in $\exp(-2K)$. At large $K$ the leading
term in entropy is $h(K)\simeq \hbox{$(2K+1)$}\exp(-2K)$ hence Eq. (\ref{ii}) results
in
\begin{equation}\label{eq}
  (2K+1-\ln2)\exp(-2K)=(K'+\frac{1}{2})\exp(-2K'),
\end{equation}
which has the solution
\begin{equation}\label{k}
  K'=K-\frac{1}{2}\ln2+\ldots.
\end{equation}
This is the correct expression for the renormalized coupling at $K\rightarrow\infty$
and we see now that $\ln2$ here is in fact of informational origin.

Now we discuss the relative entropy Eq. (\ref{ra}) of the Ising model at some coupling
$K_1$ taken with respect to the state at a different coupling $K_2$. The general
formula, derived in \cite{G1} states that
\begin{equation}\label{rei}
  H_{rel}=W(K_1)-W(K_2)-(K_1-K_2)\frac{\partial W(K_1)}
  {\partial  K_1},
\end{equation}
where
\begin{equation}\label{w}
  W=-\ln Z.
\end{equation}
In our case $W(K)=-N\ln(2\cosh K)$ and hence
\begin{equation}\label{rei1}
  H_{rel}(K_1,K_2)=-N\ln\frac{\cosh K_1}{\cosh K_2}+N(K_1-K_2)\tanh K_1.
\end{equation}
We can look now at some limiting cases. First consider the entropy relative to the
high temperature fixed point $K_2=0$. Then
\begin{equation}\label{re0}
  H_{rel}(K,0)=N(\ln 2-h(K))=N\ln 2-H.
\end{equation}
One can easily see that $H_{rel}$ decreases under the RG transformation when
$K\rightarrow K'$ and $N\rightarrow N/2$ and $h_{rel}=H_{rel}/N$ also tends to zero.

The entropy relative to the zero temperature fixed point should be obtained in the
limit $K_2\rightarrow\infty$ but the limiting expression diverges. Fortunately, we can
take $K_1=\infty$ which is almost the same for our purposes. Then
\begin{equation}\label{reinf}
  H_{rel}(\infty,K)=N[-K+\ln(2\cosh K)].
\end{equation}
Again the total relative entropy monotonically decreases along the RG flow as the
general proof suggests, but now $h_{rel}$ increases from zero to $\ln 2$. Close to the
critical point
\begin{equation}\label{rega}
H_{rel}(\infty,K)\simeq N\exp(-2K)=2N/\xi, \qquad K\rightarrow\infty
\end{equation}
where $\xi\simeq 1/2\exp(2K)$ is the correlation length. Thus, in this limit $H_{rel}$
is not changed by the RG transformation $\xi\rightarrow\xi/2$, $N\rightarrow N/2$ and
actually measures the number of blocks of size $\xi$ as the general analysis suggests
\cite{G1,G2}.

This example clearly illustrates that $H_{rel}$ always decreases, but this behavior is
in part due to the effective shrinking of the system $N\rightarrow N/2$ and implies
only that $h_{rel}(K')\leq 2h_{rel}(K)$. The relative entropy per lattice site
$h_{rel}$ is more useful because it really measures the ``distance'' from the fixed
points and its evolution is related to the direction of the flow. Since here the RG
flow is from the zero temperature fixed point to the high temperature one,
$h_{rel}(K,0)$ decreases while $h_{rel}(\infty,K)$ monotonically grows (as $1/\xi$
close to the critical point).

It is easy also to calculate the mutual information $I$ for this model. Since the two
sublattices are identical and similar to the original one,
$H(\psi)=H(\eta)=H'=(N/2)h(K')$ and hence
\begin{equation}\label{imu}
  I=2H'-H=N[h(K')-h(K)]
\end{equation}
We see then, that indeed $I$, as expected, is determined entirely by the change of the
coupling constant and from $I\geq 0$ it follows that
\begin{equation}\label{hgr}
h(K')\geq h(K),
\end{equation}
i.e. the entropy per site $h(K)$ always grows along the RG trajectory. This is rather
trivial for the model in question since the RG flow leads us to the disordered fixed
point $K=0$ with maximum entropy regardless of the initial state because there is no
phase transition. Nevertheless this simple example clearly shows how the
irreversibility of the RG flow can be derived within the informational approach.

So far we have considered only the thermodynamic limit $N\rightarrow\infty$. Let us
now take a more careful look at the finite system. The partition function for finite
block of Ising spins is well known $Z=2(2\cosh K)^{N-1}$. From this expression it is
easy to obtain the entropy and its expansion in powers of $1/N$
\begin{equation}\label{exp}
  H(N)=Nh+C+\ldots,
\end{equation}
where $h$ is given by Eq. (\ref{h}) and
\begin{equation}\label{cis}
  C=K\tanh K-\ln (\cosh K).
\end{equation}
This $C$ monotonically decreases from $\ln 2$ at $K=\infty$ to zero at $K=0$. It is
interesting, that at fixed points $C$ is equal to universal values $\ln q$, where $q$
is the degeneracy of the largest eigenvalue of the transfer matrix
$T_{s,s'}=\exp(Kss')$, $s,s'=\pm 1$. Indeed, $q=2$ at zero temperature fixed point
(due to two ground states with opposite magnetization) while $q=1$ in the fully
disordered state.

This subextensive part of the entropy is also known as the {\em excess entropy} and
was studied in detail as a possible measure of statistical complexity of spatial
structures in 1D spin systems \cite{FC}. It can be shown that if we divide our finite
system into two halves then excess entropy $C$ is equal to the mutual information of
these halves (see \cite{FC} and references therein). But any mutual information is a
kind of relative entropy and so it should in general decrease under coarse graining
(if it preserves the partition of the block in two parts) \cite{Cov} (see also Eq.
(\ref{ca})) and since it does not depend on $N$ this decrease is not related to the
change of size $N\rightarrow N/2$. Thus for 1D Ising model the excess entropy $C$ has
an independent meaning and its monotonic decrease under renormalization can be derived
directly from the information theory.

\subsection{Classical spin models with $s\geq 1$ in one dimension}

More interesting critical behavior may be obtained in one dimension if we take a spin
chain with larger value of spin $s$. We start with the most studied example of a
spin-1 chain which is defined by the action
\begin{equation}\label{spoH}
  S=-K_1\sum_{i}s_is_{i+1}-K_2\sum_{i}s^2_is^2_{i+1}+K_3\sum_i s^2_i,
\end{equation}
where spin now takes three possible values $s_i=0,\pm 1$. This is again a
ferromagnetic model for $K_1>0$, but with much richer behavior than before because we
can now vary the effective length of the spin by changing the parameter $K_3$. The
action (\ref{spoH}) retains its form under the decimation transformation and only the
coupling constants are changed.

It is widely believed that there are no phase transitions in one-dimensional systems
with short range interactions. This is true, however, only if all the couplings are
finite. Non-trivial critical behavior in this model, analyzed thoroughly in Ref.
\cite{spo}, takes place when both $K_2$ and $K_3$ tend to infinity while $K_2-K_3$
remains finite. The remaining two-dimensional space of couplings may be parameterized
by new coordinates
\begin{equation}\label{xy}
  x=\exp(-K_1-K_2+K_3),\qquad y=\exp(-2K_1)
\end{equation}
and the decimation transformation acts as
\begin{equation}\label{spoRG}
  x'=\frac{x^2}{1+y^2}, \quad y'=\frac{2y}{1+y^2}.
\end{equation}
\begin{figure}[h]
\includegraphics[width=3in]{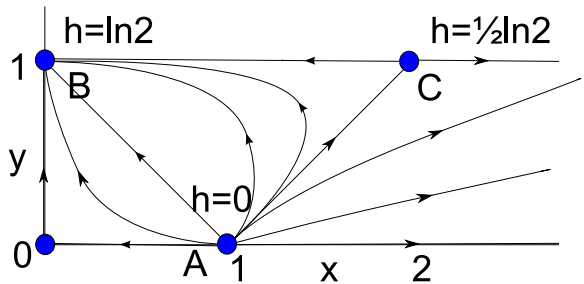}
\caption{Decimation RG flow for the spin-1 model \cite{spo} with the values of entropy
per lattice site $h$ at some fixed points.} \label{fig2}
\end{figure}

The resulting quite interesting RG flow diagram obtained in \cite{spo} is shown
schematically in Fig. 2. Its main features are four fixed points and a critical line
AC. Since this flow arises from the decimation transformation our general arguments
require that the entropy per site $h(x,y)$ should increase along the flow. Though the
model can be solved exactly (see \cite{spo} for details) it is easy to evaluate the
entropy at the fixed points without much computations.

The point A corresponds to fully ordered state (actually there are three such states
with all $s_i$ either $+1$, or $-1$, or 0) hence $h_A=H/N=0$ at $N\rightarrow\infty$
with only subextensive contribution to $H$. The same is true also for the point
$x=y=0$ with two ordered states. The point B is similar to a high temperature fixed
point of the Ising model, with uncorrelated $s_i=\pm 1$ and no zero spins, hence
$h_B=\ln2$.

The most interesting case is the critical point C at $x=2,y=1$. It corresponds to
$K_1=0$ while $K_3-K_2=\ln2$. In this case there exist the lowest energy state with
all $s_i=0$ and zero energy. For periodic boundary condition all configurations with
some finite fraction of spins changed to $\pm 1$ have infinite energy at
$K_3\rightarrow\infty$ while states with all $s_i=\pm 1$ are gapped with $\Delta
E=N\ln2$. But there are $2^N=\exp(N\ln2)$ such states and their statistical weight
exactly compensates the Boltzmann factor $\exp(-\Delta E)$ which results in the
partition function $Z=2$ and mean energy $\langle S\rangle=\Delta E/2$. Therefore at C
we effectively have two equally probable states with highly degenerate disordered
excited one and the resulting entropy $h_C=(\langle S\rangle+\ln Z)/N\rightarrow
(1/2)\ln2$ at $N\rightarrow\infty$.

The values of entropies are shown in Fig. 2 near the corresponding fixed points. We
see now that the entropy per site can in fact be used to understand the ordering of
the fixed points along the flow since $h$ monotonically grows when we pass from A to C
and finally to B.

Certainly this is not the only way to classify these fixed points. In fact we can use
also the degeneracy $q$ of the largest eigenvalue of the transfer matrix $T$ (actually
the degeneracy of the ground state of the corresponding quantum Hamiltonian $\hat{H}$,
defined by $\hat{T}=\exp(-\hat{H})$). In 1D $q$ plays, in a sense, the same role as
the central charge $c$ in two dimensions, because $q$ always decreases under
renormalization (this also resembles the $g$-theorem \cite{g} for 1D quantum systems).
The global decrease of $q$ in 1D spin system with nearest-neighbor interactions when
\begin{equation}\label{near}
  S=\sum_{i}V(s_i,s_{i+1})
\end{equation}
is rather trivial and follows directly from the decimation RG transformation written
as a transfer matrix mapping
\begin{equation}\label{tren}
  T\rightarrow T'=T^2, \qquad T_{s,s'}=\exp(-V(s,s'))
\end{equation}
where for spin with $n$ possible values $T$ is a nonnegative $n\times n$ matrix.
Eigenvalues of $T$ transform as $\lambda_i\rightarrow\lambda_i '=\lambda_i^2$ so that
the degeneracy once reduced will never return back. For strictly positive $T$
Frobenius theorem requires $q=1$, but when some couplings turn to infinity, as we have
seen above for the spin-1 model, some entries in $T$ may vanish and $q$ may differ
from unity. For the spin-1 model all these eigenvalues were calculated in Ref.
\cite{spo} and $q=3$ for point A, $q=2$ for C (also for the point at the origin) and
$q=1$ for point B in Fig. 2.

For these models it is possible also to evaluate the excess entropy at the fixed
points. Fixed points for decimation are defined by
\begin{equation}\label{tfix}
T'=T^2=\lambda T
\end{equation}
where $\lambda$ is some positive number (multiplication of $T$ by a constant does not
change the physics) and it can be easily shown that in these cases $T$ may have $q\leq
n$ eigenvalues $\lambda$ while other eigenvalues are zeros. Then for the periodic
boundary condition
\begin{equation}\label{spn}
  Z={\rm tr}(T^N)=q\lambda^N,\quad \Rightarrow\quad\ln Z=N\ln\lambda+\ln q
\end{equation}
and since $\langle S\rangle\sim N$ entropy at fixed points has the form $Nh+C$ with
$C=\ln q$. In this case the subextensive part in $H$ coincides with that in $\ln Z$
and this number monotonically decreases when we pass from one fixed point to another
along the RG flow (this is somewhat similar in spirit to the recently proposed
$F$-theorem for field theories in odd dimensions \cite{Ftheor}). Unfortunately, for
periodic boundary conditions $C$ equals $\ln q$ also away from fixed points (at
$N\rightarrow\infty$) and it is discontinuous, because $q$ changes abruptly, while
excess entropy for a finite block in an infinite chain, which is related to mutual
information of two halves and is expected to be continuous and monotonic along the RG
flow, is much harder to evaluate at $s\geq 1$ and in general is no more related to $q$
in a simple way even at fixed points.

\subsection{Decimation in two dimensions.}

Consider now a two-dimensional classical spin system, with spins placed on sites of a
square lattice. Various spin interactions, including e.g. nearest-neighbor,
second-nearest-neighbor and similar interactions, are parameterized by a set of
coupling constants ${\bf K}=(K_1,K_2,\ldots)$. To define a ``checkerboard'' decimation
transformation we first divide the lattice into two interpenetrating square
sub-lattices and then sum over all spins on one sublattice \cite{KH} (see Fig. 1b).
Just as in one dimension we obtain then a new spin system again on a square lattice
(with the lattice spacing $\sqrt 2$ times larger) but with a new energy $S'$ with
couplings ${\bf K}'$. This is the first step of the RG transformation, which is then
repeated further.

Since two sub-lattices are identical the mutual information between them is again
given by the Eq. (\ref{imu}), i.e.
\begin{equation}\label{imu2}
  I/N=h({\bf K}')-h({\bf K})\geq 0
\end{equation}
for a large lattice with $N$ sites, where $h({\bf K})$ is the entropy per site. Hence
we conclude that in this case the entropy per site should also increase in the course
of the RG transformation and the RG flow is irreversible.

The entropy growth in two dimensions may seem a rather counterintuitive result, since
in systems with phase transition one may na\"{i}vely expect that at low temperatures
(below the transition) the RG transformation should lead us deeply into the ordered
phase with $h({\bf K}')$ decreasing to zero. But for the decimation this intuition
obviously doesn't work, since in this case the average spin remains unchanged  so that
the spontaneous magnetization is constant along the RG flow.

For better understanding of why the entropy per site can grow under renormalization at
low temperatures let us now consider the ferromagnetic Ising model with
$S=-K_{1}\sum_{{\rm nn}} s_{i}s_{j}$, where the sum is over nearest-neighbors and
$s_i=\pm 1$. After the first step of the RG we have (up to an unimportant constant)
\cite{W}
\begin{equation}\label{ieff}
  S'=-K_{1}'\sum_{{\rm nn}} s_{i}s_{j}-K_{2}'\sum_{{\rm nnn}} s_{i}s_{j}
  -K_{3}'\sum_{{\rm P}} \prod_{i=1}^{4} s_{i},
\end{equation}
where the second term is the next-nearest-neighbor interaction and the last one is the
four spin interaction ($s_i$, $i=1\ldots 4$ are spins on the corners of a plaquette)
with the sum over all plaquettes. The new coupling constants are known to be
\begin{eqnarray}\label{cc}
  & K_{1}'=2K_{2}'=\frac{1}{4}\ln(\cosh 4K_{1}),\\
  & K_{3}'=\frac{1}{8}\ln(\cosh 4K_1)-\frac{1}{2}\ln(\cosh 2K_1)
\end{eqnarray}
We see now that $K_{1}'>0$ (though $K_{1}'<K_1$), the next-nearest-neighbor
interaction is also ferromagnetic, $K_{2}'>0$, but $K_{3}'<0$, i.e. the four-spin
interaction tries to decrease the ferromagnetic order.

Next, we estimate the entropy at low temperatures. At $K_{1}\rightarrow \infty$ all
spins point in one direction and the entropy is different from zero only due to rare
spin flips. The partition function may be represented as a low temperature expansion
\begin{equation}\label{ltp}
  Z=2{\rm e}^{-NE_0}\left(1+N\exp(-\Delta E({\bf K}))+\ldots\right),
\end{equation}
where $NE_0$ is the energy of all spins aligned and the energy needed to reverse one
spin is $\Delta E=8K_1$ in original system since after the reversal four links have
``wrong'' alignment of adjacent spins. Hence at low temperatures the entropy per site
is
\begin{equation}\label{eps}
h\simeq 8K_1\exp(-8K_1)+\ldots
\end{equation}
After the decimation the system is still ferromagnetic but the energy of spin reversal
is now $\Delta E'=8K_{1}'+8K_{2}'+8K_{3}'$ or
\begin{equation}\label{de}
  \Delta E'=4\left(\ln(\cosh 4K_1)-\ln(\cosh 2K_1)\right)\simeq
  8K_1-4{\rm e}^{-4K_1}
\end{equation}
at large $K_1$. Hence the new excitation energy is smaller than $\Delta E$ and it
becomes easier to reverse a spin. However, since the difference is exponentially small
one has to include also the next term in $h({\bf K})$ due to pairs of reversed spins
$24K_1\exp(-12K_1)$ (the corresponding term in $h({\bf K}')$ can be easily evaluated
as $\sim\exp(-16K_1)$ and is unimportant). But even with this contribution taken into
account the entropy per site after the decimation $h({\bf K}')\simeq\Delta E'
\exp(-\Delta E')$ is larger,
\begin{equation}\label{e}
  h({\bf K}')-h({\bf K})  \simeq 8K_1{\rm e}^{-12K_1}>0.
\end{equation}
Thus, it is possible to check at least for the first iteration at low temperatures
that indeed the entropy per site grows after the decimation. Physically the result
(\ref {e}) for the mutual information or $I\sim N\exp(-12K_1)$ is due to the reversed
pairs of adjacent spins in the original lattice. Only through these pairs the two
sub-lattices have non-trivial information about each other at $K_1\rightarrow\infty$
(there should be also a term $\ln 2$ in $I$ since the orientations of the overall
magnetization in sub-lattices are correlated, but we are interested only in terms
$\sim N$ in the mutual information).

Now it is possible to explain, what this increase in entropy at low temperatures
actually means. Since the average magnetization conserves along the decimation RG
flow, the average number of reversed spins per unit volume remains the same, but now
more of them are isolated ones (with less energy cost of reversal) and the number of
reversed close pairs or larger clusters is relatively smaller than in the initial
system. Surely this picture may also be interpreted as a decrease in the correlation
length for spin fluctuations.

Unfortunately the next iteration of the decimation cannot be performed exactly and is
known to lead to non-Gibbsian measures at low temperatures \cite{nG}, so that it is
not possible to describe renormalization as a flow in usual space of couplings. Still
there should be a flow of probability measures, probably with limiting points
corresponding to product measures describing independent spins with fixed total
magnetization.

\subsection{Kadanoff block-spin transformation}

Consider now a classical lattice system of Ising spins $s_i$ with the action $S_0[s]$
and first divide the lattice into a set of nonoverlapping identical blocks with, say,
$k$ adjacent spins in each block $B_j$, assuming that blocks form a lattice, similar
to the original one (see Fig. 3 for an example). Then, following Kadanoff prescription
\cite{K} associate a new spin variable $t_j$ with $j$-th block and define the RG
transformation $S_0[s]\rightarrow S'[t]$ according to
\begin{equation}\label{K}
  \exp(S'[t])=\sum_{s}T(t,s)\exp(-S_0[s])
\end{equation}
with $T(t,s)>0$ obeying the condition
\begin{equation}\label{T}
  \sum_t T(t,s)=1
\end{equation}
which guarantees that the partition function $Z$ is the same for both $s$ and $t$
variables.

\begin{figure}[h]
\includegraphics[width=2.5in]{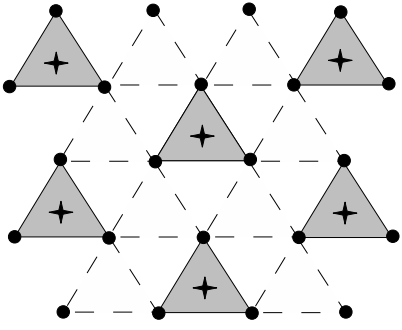}
\caption{Block-spin transformation on a triangular lattice for $k=3$. Blocks $B_j$ are
shaded, spins $s_i$ are located at black sites while block variables $t_j$ may be
placed in the centers of blocks.} \label{fig3}
\end{figure}

This transformation is of the general type of Eq. (\ref{eff}) but now all $s_i$ are
``fast'' variables and $t_j$ are ``slow'' ones. Hence the total ``action''
$S[t,s]=-\ln T(t,s)+S_0[s]$ now defines the joint probability for initial and
``renormalized'' spins. The entropy $H$ now is not equal to the physical entropy $H_0$
of the initial spin system but is given by
\begin{eqnarray}\label{HK}
  & H(t,s)=\langle S\rangle_S+\ln Z= \langle
  S_0\rangle_{S_0}-\langle\ln T\rangle_S+\ln Z=\\ \nonumber
  & =H_0-\langle\ln T\rangle_S,
\end{eqnarray}
where we have used the condition (\ref{T}) in the calculation of $\langle
S_0\rangle_S$.

The most popular form of $T(t,s)$ is
\begin{equation}\label{Tp}
  T(t,s)=\prod_j \frac{\exp(pt_jS_j)}{2\cosh(pS_j)},
\end{equation}
where $p$ is a free parameter and
\begin{equation}\label{Ss}
  S_j=\sum_{i\in B_j}s_i
\end{equation}
is the sum of initial spins in the block $B_j$. Note that $pt_j$ enters in the
partition function as a spatially varying magnetic field constant over a block. Hence
at $p\rightarrow\infty$ all block spins $S_j$ will be aligned strictly along $t_j$.
Now making use of the identity $\langle t_j\rangle_{S}=\langle
\tanh(pS_j)\rangle_{S_0}$ we easily obtain
\begin{eqnarray}\label{Tav}
& -\langle\ln T\rangle_S=\sum_j \langle f(pS_j)\rangle_{S_0},\,\\
\nonumber & f(x)=-x\tanh x+\ln(2\cosh x),
\end{eqnarray}
where the sum is over all blocks. Next, we can easily sum $\exp(-S[t,s])$ over $t$
because of the identity (\ref{T}) to obtain $\tilde{S}'[s]=S_0[s]$ and hence
\begin{equation}\label{Ktild}
  H(s)=H_0
\end{equation}

Now we can use the definition of the mutual information (\ref{mu}) and Eqs.
(\ref{HK}), (\ref{Ktild}) to derive the final expression
\begin{equation}\label{muK}
  I=H(t)+H(s)-H(t,s)=H(t)-\sum_j \langle f(pS_j)\rangle_{S_0}
\end{equation}
Note that $H_0$ cancels away from the resulting expression for the mutual information.
Next we introduce the entropy per lattice site for renormalized spins according to
$H(t)=(N/k)h({\bf K}')$ and instead of $I=N(h({\bf K}')-h({\bf K}))$ which was valid
for decimation, we have at large $N$ quite a different expression
\begin{equation}\label{muBS}
  I=\frac{N}{k}[h({\bf K}')-\langle f(p\sum_i s_i)\rangle_0]\geq 0,
\end{equation}
where the sum of spins $s_i$ is over one block, $k$ is the number of spins in a block
and the average is over the initial configurations with the weight $\exp(-S_0[s])$.
The second term in Eq. (\ref{muBS}) looks like the entropy of the total spin of a
block at temperature $1/p$ averaged over different values of its magnitude.

Thus the entropy after the renormalization $h({\bf K}')$ is always bounded from below,
provided the second term in Eq. (\ref{muBS}) is nonzero. This is due to the noise
induced by the renormalization at finite $p$.

The most interesting case however corresponds to $p\rightarrow\infty$ which is known
to lead to the so called majority-rule renormalization \cite{MR}. This means that we
define $t_j$ as $\pm 1$ depending on the sign of the total block spin $S_j$. We assume
also that $k$ is odd to avoid ambiguities. Then $S_j\neq 0$ and from Eq. (\ref{Tav})
it follows that $f(pS_j)\rightarrow 0$ at $p\rightarrow\infty$ leading to
\begin{equation}\label{mutwo}
  I=H'=\frac{N}{k}h({\bf K}'), \qquad p\rightarrow\infty, \qquad k-{\rm
  odd}.
\end{equation}
This result is rather obvious however, because from Eq. (\ref{mu}) we have
$I=H'-H(t|s)$ but for majority-rule RG the initial configuration of spins $s_i$
completely determines that of renormalized spins $t_j$ so that $H(t|s)=0$ (there is no
noise in the transmission channel) and hence their mutual information coincides with
the final entropy $I=H'$. Therefore in this case the positivity of mutual information
gives nothing new and $h({\bf K}')$ can be either greater or smaller than $h({\bf
K})$. Contrary to decimation now $I\geq 0$ does not prohibit fixed points and the
entropy per site can even go to zero along the RG flow.

\subsection{One dimensional free field theory}

Now we proceed to quantum field theory, using ordinary quantum mechanics as the
simplest example. Consider a quantum particle of mass $m$, moving along a line in a
harmonic potential well with a frequency $\omega$. Its partition function at a
temperature $T$ is given by a path integral, which in a lattice regularization, when
the imaginary time interval $[0,\beta]$ ($\beta=1/T$) is divided into $N$ intervals of
length $\epsilon$, has the form
\begin{eqnarray}\label{zA}
  Z=\left(\frac{m}{2\pi\epsilon}\right)^{N/2}\int\exp(-S)\prod_{i=1}^N
  dx_i, \nonumber\\
  S=\sum_i \frac{m}{2}\frac{(x_{i+1}-x_{i})^2}{\epsilon}+
  \epsilon \sum_i \frac{m\omega^2}{2}x_i^2,
\end{eqnarray}
with $N=\beta/\epsilon$. Shannon entropy, as mentioned above, equals
\begin{equation}\label{hA}
  H=\langle S\rangle+\ln Z,
\end{equation}
where brackets mean the average with weight $\exp(-S)$. $H$ coincides with the entropy
of the statistical system of variables $x_i$ placed on sites of a one-dimensional
lattice of length $\beta$ with nearest-neighbor interaction and periodic boundary
condition. This system is actually the one-dimensional Gaussian model near its
critical point $\omega=0$ \cite{GA}.

However, at $\epsilon\rightarrow 0$ this model is actually (0+1) Euclidean free field
theory and its thermodynamic entropy is given by a different formula
\begin{equation}\label{hq}
  H_T=(E-F)/T=E\beta+\ln Z
\end{equation}
where $E$ is the mean energy of the particle and $F=-T\ln Z$ is its free energy. Let
us now briefly discuss the connection between these two entropies.

The mean velocity squared $\langle\dot{x}^2\rangle$ which enters in $\langle S\rangle$
diverges as $\epsilon\rightarrow 0$ \cite{F} and its direct evaluation is not so easy,
but the average action can be evaluated simply by differentiating $Z$ with respect to
mass
\begin{equation}\label{z1A}
  m\frac{\partial Z}{\partial m}=\frac{N}{2}Z-\langle S\rangle Z,
\end{equation}
where the first term in the r.h.s. comes from the differentiation of $m$ in the
integration measure. But for the harmonic oscillator $Z$ does not depend on mass, and
therefore $\langle S\rangle =N/2$. In the continuum limit
\begin{equation}\label{HohA}
  \ln Z=-\ln(2\sinh(\omega\beta/2)),
\end{equation}
and at a given $\omega$  we have the following asymptotics for the entropies at
$\beta\rightarrow\infty$
\begin{eqnarray}\label{HA}
 H=\frac{\beta}{2}\left(\frac{1}{\epsilon}-\omega\right)+
 {\rm e}^{-\omega\beta}+\ldots\\
H_T=\omega\beta{\rm e}^{-\omega\beta}+\ldots
\end{eqnarray}
We see that the leading ``bulk'' term in $H$ is divergent in the continuum limit
$\epsilon\rightarrow 0$ and $H_T$ resembles, in a sense, the finite size correction in
$H$.

The different behavior of $H$ and $H_T$ comes from the fact that $H_T$ describes the
information related to energy levels occupation and hence tends to zero at zero
temperature when the system is in its ground state, while $H$ describes fluctuations
of ``paths'' in configuration space which obviously grow as $\beta\rightarrow \infty$.

The entropy per lattice site here is
\begin{equation}\label{hf}
  h=\frac{1}{2}-\frac{\epsilon\omega}{2}+\ldots
\end{equation}
at $\epsilon\rightarrow 0$. It may seem strange that $h$ decreases with correlation
length $\xi=(\omega\epsilon)^{-1}$, because for the model in question we may also
apply the decimation transformation, similar to that in the Ising model, integrating
away $x_i$'s on alternating sites (see e.g. \cite{Kb}) and our general result states
that in this case $h$ should increase along the RG flow. Under such a decimation
$\xi\rightarrow\xi/2$ and $h$ from Eq. (\ref{hf}) does not behave in the required way.

The solution to this apparent contradiction is rather simple. Note first that
summation over variables $x$ is defined now as
\begin{equation}\label{sum}
  \sum_{x}(\ldots)=\int\prod_i \left(\sqrt{\frac{m}{2\pi\epsilon}}
    dx_i\right)(\ldots)
\end{equation}
and the integration measure itself depends on mass and on lattice spacing. This
integration measure is fine-tuned to the action so that $Z$ has a well defined correct
continuum limit (\ref{HohA}). But after the RG transformation this fine-tuning is
destroyed. The effective action $S'$ now contains renormalized mass $m'$, frequency
$\omega'$ and a new lattice spacing $2\epsilon$, but the measure (\ref{sum}) remains
the same. This means that in this case {\em the entropy after renormalization is no
longer given by the same formula (\ref{HA}) as before}.

We may transform the measure to the correct form but this leads to an additional
factor $(2m/m')^{N/4}$ in $Z$, since the lattice now has $N/2$ sites. This additional
factor does not affect the mean action $\langle S'\rangle$ but certainly change the
$\ln Z$ part of the entropy. Simple Gaussian integration like that in Ref. \cite{Kb}
shows that $m'\simeq m$ up to the terms of order $\epsilon^2$ at $\epsilon\rightarrow
0$ and the additional factor is simply $2^{N/4}$. Then the new entropy per site equals
\begin{equation}\label{hpr}
  h'=\frac{1}{2}+\frac{1}{2}\ln2+\ldots,
\end{equation}
at $\epsilon\rightarrow 0$. This quantity is larger than $h\simeq 1/2$ and the mutual
information $I$ of two sublattices is given by
\begin{equation}\label{muf}
  I= N(h'-h)=\frac{N}{2}\ln2
\end{equation}
at $\epsilon\rightarrow 0$. Note also that the total entropy is reduced after the
decimation, because the information loss $\delta I=H-H'=N(1-\ln 2)/4$ is positive.

Let us now analyze this system more carefully. After the decimation is repeated many
times we get away from the continuum limit because the lattice spacing grows. In this
region it is more convenient to rewrite the renormalized action in the form
\begin{equation}\label{sren}
  S=\sum_{i}Kx_{i}x_{i+1}+\frac{b}{2}\sum_{i}x_{i}^2,
\end{equation}
where $K$, $b$ are new running coupling constants with the initial values
$K_0=m/\epsilon$, $b_0=2m/\epsilon+\epsilon m\omega^2$. The entropy per lattice site
for this action can be derived from the exact solution of the Gaussian model \cite{GA}
\begin{eqnarray}\label{HGA}
  h&=&\frac{1}{2}+\frac{1}{2}\ln\left(\frac{K_0}{b}\right)-
  \frac{1}{2}\int_0^{2\pi}\frac{dt}{2\pi}\ln\left(1-2\frac{K}{b}\cos t\right)
  = \nonumber\\
  &&=\frac{1}{2}+\frac{1}{2}\ln\left(z\frac{K_0}{K}\right)-\frac{1}{2}
  \ln{\frac{1+\sqrt{1-4z^2}}{2}}, \quad z=\frac{K}{b},
\end{eqnarray}
where $K_0=m/\epsilon$ comes from the integration measure in Eq. (\ref{zA}). Note that
while all correlations depend only on $z=K/b$, the entropy depends on both $K$ and $b$
because the integration measure contains the unrenormalized coupling $K_0$. Decimation
acts as
\begin{equation}\label{dec}
  K'=\frac{K^2}{b},\quad b'=b-2\frac{K^2}{b},
\end{equation}
or $z'=z^2/(1-2z^2)$. This RG equation has a fixed point $z^*=1/2$ which corresponds
to $\omega=0$ and under renormalization $z$ flows to $z=0$. Close to the fixed point
$z=1/2$ one can define a continuous theory by choosing $K\simeq b/2\simeq m/\epsilon$
and $b-2K=\epsilon m\omega^2$ with $\epsilon\rightarrow 0$ and RG equations turn into
$K'\simeq K/2$, $b'\simeq b/2$.

It is now easy to verify that in the vicinity of $z=1/2$ the initial value of the
entropy is $h=1/2$ and that it indeed acquires an additional $(1/2)\ln2$ contribution
at each decimation step provided we are still close to the fixed point. Note that $h$
grows even for $z=1/2$ because in the two dimensional space of $K$ and $b$ there is no
critical fixed point and the point $z=1/2$ corresponds to a line $K=b/2$ with the RG
flow along it. After many decimations $z\rightarrow 0$, i.e. $K$ decreases much faster
than $b$ and the entropy starts to behave as
\begin{equation}\label{lim}
  h\simeq\frac{1}{2}\ln\left(\frac{K_0}{b}\right),
  \quad K\ll b\ll K_0
\end{equation}
with logarithmic accuracy. Since $K\rightarrow 0$ means the loss of correlations
between $x$'s on different sites, this expression for the entropy $h$ is in fact the
entropy production rate for the Gaussian white noise $h\sim\ln\sigma$ with dispersion
$\sigma^2\sim 1/b$, well known in the information theory.

\section{Conclusion}

In this paper we have tried to understand what limitations on the RG flow may follow
from the information theory, because Wilsonian RG transformation, when some fast
variables are integrated out, is obviously related to the information loss. This
transformation is similar in a sense to a signal transmission through a noisy channel,
the fast variables being the input, while the slow ones play the role of the output
available to a receiver. The information loss is defined here as an average
conditional entropy of the eliminated fast variables when the remaining slow variables
are fixed. From the nonnegativity of this quantity (which holds at least for systems
with discrete variables, like e.g. classical spins) it follows immediately that the
informational entropy $H$ of the system cannot increase under renormalization.

Unfortunately this result is not very interesting, and does not lead to the
irreversibility of the RG flow, because it applies to the total extensive entropy,
proportional to the size of the system, which effectively shrinks after
renormalization. This shrinking is evident for real space RG schemes, where some
lattice sites are eliminated and the system size is smaller if measured in new lattice
constant, but equally applies to momentum space Wilsonian renormalization due to final
rescaling of momenta.

Only if $H$ possesses some volume independent additive part $C$, then this part, if it
also decreases under renormalization, may provide the required Lyapunov function, that
depends only on couplings. Physically its decrease may be attributed to a loss of
information about massless modes, but this does not follow directly from that of $H$
and should be deduced independently. Then, to be monotonous $C$ must have also some
independent definition. For 1D Ising model this quantity is just the subextensive
excess entropy and we explicitly check its monotonicity under RG transformation. This
behavior for a finite block of spins in an infinite chain follows directly from
information theory, because excess entropy in this case is just the mutual information
of two halves of the system \cite{FC}, which decreases under coarse graining.

Note also that for 1D systems with periodic boundary conditions the excess entropy is
$\ln q$ where $q$ is the degeneracy of the largest eigenvalue of the transfer matrix
which always decreases along the RG flow. This is somewhat similar to what is known
for critical 2D systems defined on a long cylinder of size $L_x\times L_y$ with, say,
$L_y\rightarrow\infty$ and periodic boundary condition in the $x$ direction. In this
case the first finite size correction in the expansion of entropy in $1/L_x$ is $\sim
c(L_y/L_x$) \cite{fs}, where $c$ is the central charge. Therefore the subextensive
part $C$ in entropy in 2D should decrease as a consequence of the $c$-theorem. However
in both these examples the subleading term in entropy is discontinuous and does not
directly lead to a proper Lyapunov function.

This means that the extraction of the required part from the entropy of finite size
system is somewhat ambiguous and depends on the boundary condition. One may suggest,
however, that in general there might exist some size-independent information-based
quantity defined on the whole space of couplings (maybe some kind of relative entropy
or higher order mutual information \cite{ho}) that is monotonic along the flow and at
fixed points coincides with some kind of excess entropy, but this point is not clear.

The main aim of the present paper was, however, apart from demonstrating some examples
of informational approach, to introduce a different quantity, namely the mutual
information of fast and slow variables $I$, which shows how much information about the
eliminated variables is still stored in the renormalized effective action. This
quantity does not suffer from the ``shrinking size'' trouble and is less sensitive to
the overall decrease in the number of degrees of freedom, though it is generally more
difficult to evaluate.

There exist however some real space decimation transformations for which this mutual
information can be easily calculated. This happens e.g. if the original lattice may be
decomposed in two identical sublattices and RG transformation eliminates spins on one
sublattice. Then the nonnegativity of $I$ results in the increase of entropy per
lattice cite and RG flow resembles a relaxation to equilibrium. This result does not
depend on interactions and may be true also for a larger set of models. What is
actually needed is that the RG transformation must be represented as a product of such
decimations.

The approach to irreversibility based on the mutual information is different from
those trying to generalize Zamolodchikov's $c$-theorem and possibly may provide some
additional information about the RG flows. It may appear interesting also to apply the
present approach to the momentum space renormalization in field theory or to study
more complicated correlations along the flow and some higher-order mutual information.

I am very grateful to A.C.D. van Enter, J. Gaite, V. Losyakov, A. Marshakov, A.
Morozov for valuable discussions and comments, and to L. Shchur for some useful
advices. The work was supported  in part by RFBR grants 09-02-00886,  10-02-00509,
RFBR-Ukraine 09-01-90493 and by Federal Agency for Science and Innovations of Russian
Federation under contract 14.740.11.0081.

\end{document}